\documentclass[conference]{IEEEtran}
\usepackage{amsmath, amssymb, bm, cite, epsfig, psfrag}
\usepackage{graphicx}
\usepackage{soul} 
\usepackage[left=0.625in, right=0.625in, top=0.7in, bottom=0.95in]{geometry}
\usepackage{dblfloatfix}
\usepackage{array}
\usepackage{lipsum}
\usepackage{subcaption}

\newcolumntype{P}[1]{>{\centering\hspace{0pt}}p{#1}}
\newcolumntype{M}[1]{>{\centering\hspace{0pt}}m{#1}}
\newcolumntype{L}{>{\centering\arraybackslash}m{3cm}}
\usepackage{caption}
\usepackage{epstopdf}	
\usepackage{longtable}
\usepackage{supertabular,booktabs}
\usepackage{bbm}
\usepackage{multirow}
\usepackage[usenames,dvipsnames]{xcolor}

\usepackage{etoolbox}
\usepackage{pbox}
\usepackage{fixltx2e}
\usepackage{tabu}	
\usepackage{filecontents}
\usepackage{enumerate}
\usepackage{textcomp}
\usepackage{makecell}
\usepackage{gensymb}
\usepackage{colortbl}
\usepackage{fancyhdr}
\def\PL{\mathrm{PL}}

\newtoggle{conference}

\togglefalse{conference} 
\graphicspath{{figures/}}

\newcolumntype{?}{!{\vrule width 2pt}}

\fancypagestyle{firststyle}{
	\fancyhf{}
	\fancyhead[L]{S. Ju, Y. Xing, O. Kanhere and T. S. Rappaport, ``Sub-Terahertz Channel Measurements and Characterization in a Factory Building,''  2022 \textit{IEEE International Conference on Communications (ICC)}, May 2022, pp. 1-6.}     

}
\IEEEoverridecommandlockouts

\begin{document}
\title{Sub-Terahertz Channel Measurements and Characterization in a Factory Building} 
\author{\IEEEauthorblockN{Shihao Ju, Yunchou Xing, Ojas Kanhere, and Theodore S. Rappaport}

\IEEEauthorblockA{	\small NYU WIRELESS, Tandon School of Engineering, New York University, Brooklyn, NY, 11201\\
				\{shao, ychou, ojask, tsr\}@nyu.edu}
					\thanks{This research is supported by the NYU WIRELESS Industrial Affiliates Program and National Science Foundation (NSF) Research Grants: 1909206 and 2037845.}
}

\maketitle
\thispagestyle{firststyle}
\begin{abstract}
Sub-Terahertz (THz) frequencies between 100 GHz and 300 GHz are being considered as a key enabler for the sixth-generation (6G) wireless communications due to the vast amounts of unused spectrum. The 3rd Generation Partnership Project (3GPP) included the indoor industrial environments as a scenario of interest since Release 15. This paper presents recent sub-THz channel measurements using directional horn antennas of 27 dBi gain at 142 GHz in a factory building, which hosts equipment manufacturing startups. Directional measurements with co-polarized and cross-polarized antenna configurations were conducted over distances from 6 to 40 meters. Omnidirectional and directional path loss with two antenna polarization configurations produce the gross cross-polarization discrimination (XPD) with a mean of 27.7 dB, which suggests that dual-polarized antenna arrays can provide good multiplexing gain for sub-THz wireless systems. The measured power delay profile and power angular spectrum show the maximum root mean square (RMS) delay spread of 66.0 nanoseconds and the maximum RMS angular spread of 103.7 degrees using a 30 dB threshold, indicating the factory scenario is a rich-scattering environment due to a massive number of metal structures and objects. This work will facilitate emerging sub-THz applications such as super-resolution sensing and positioning for future smart factories.
 
\end{abstract}

\begin{IEEEkeywords}                            
Channel Measurement; Channel Modeling; Channel Statistics; Indoor Factory; Sub-Terahertz; 140 GHz; 5G; 6G 
\end{IEEEkeywords}

\section{Introduction} \label{sec:intro}
With the advent of manufacturing and information technologies, the next industry evolution, Industry 4.0, has been widely proposed and studied worldwide, aiming to escalate the production capability for future decades \cite{Chen18access}. As a crucial part of Industry 4.0, smart factory presents the digitization and intellectualization of the vertical and horizontal integration of all participants in the production process. Integrating the physical machines with the virtual software into a single cyber-physical system will drastically increase production efficiency. Furthermore, creating a digital copy of all components in a physical factory called digital twin can provide unprecedented fine control and monitoring capability of every production piece \cite{Tao19tii}. Core applications such as the industrial Internet of things (IIoT) and artificial intelligence require multiple sensors connected, data collected and processed, and finally, responses made in a timely and intelligent manner \cite{Cheng18a}. Thus, the wireless network needs to support data rates of tens of Gigabits per second (Gbps), low-latency connections of less than one millisecond, and provide high-resolution sensing and imaging capabilities. 

To fulfill such demanding requirements, moving to the sub-Terahertz (THz) spectrum (100 – 300 GHz) is paramount due to the vast available bandwidth of tens of GHz \cite{Rap19access}. The ultrawide bandwidth can be leveraged to offer an extremely high data rate and signal resolution, which leads to better wireless connectivity and sensing capability. The Federal Communications Commission (FCC) has authorized four unlicensed bands (116-122, 174.8-182, 185-190, and 244-246 GHz) in 2019 \cite{FCC19}. Since signal propagation characteristics in such a wide range of frequencies can be drastically distinct, it is essential to gain fundamental knowledge of propagation characteristics at sub-THz frequencies. Factories have been proposed as a distinct environment in the 3rd Generation Partnership Project (3GPP) Release 15 \cite{3GPP38901r16} for wireless propagation from other indoor environments such as offices and shopping malls due to the vast existence of metallic machines and robotics. Smooth metallic surfaces are more reflective and less penetrative than common building materials (e.g., drywall) for wireless signals, especially at sub-THz frequencies. Therefore, extensive channel measurements and accurate channel models are vital to establishing future sub-THz wireless networks for manufacturing factories.

Mismatched characteristics of polarization of the transmit antenna, propagation channel, and receive antenna causes degradation in the received signal power \cite{Maltsev10AwpLetter}. Polarization has been widely used at mmWave frequencies as a degree of freedom other than time, frequency, and space, to support multiple input multiple output (MIMO) diversity and multiplexing \cite{Sun14CommMag,3GPP38901r16,80211ad10}. The sub-THz channels have been observed to be more sparse than mmWave channels \cite{Ju21jsac, Xing21b}. Thus, the dual-polarized transmit and receive antenna arrays are necessary to deliver more signal energy and increase channel rank. 

The work presents the 142 GHz channel measurements conducted in a factory building using directional horn antennas with two polarization settings (i.e., vertical-to-vertical (V-V) and vertical-to-horizontal (V-H)) at two transmitter (TX) locations and seven receiver (RX) locations, aiming to understand the channel characteristics in terms of path loss, delay spread, angular spread, and polarization effect in the indoor factory (InF) scenario at sub-THz frequencies. The rest of the paper is organized as follows: Section \ref{sec:meas} describes the factory environment, measurement settings, and procedure. Section \ref{sec:pl} presents the omnidirectional and directional path loss using V-V and V-H antenna polarization and gives the gross cross-polarization discrimination (XPD) values for the omnidirectional channel and the XPD for directional paths (e.g., boresight and reflection paths). Section \ref{sec:dsas} derives the mean and maximum root mean square (RMS) delay spread and RMS angular spread across all measured locations. Finally, concluding remarks and future work are provided in Section \ref{sec:conclusion}.

\section{142 GHz InF Channel Measurements} \label{sec:meas}
\begin{figure}[h!]
\centering
\includegraphics[width=1\linewidth]{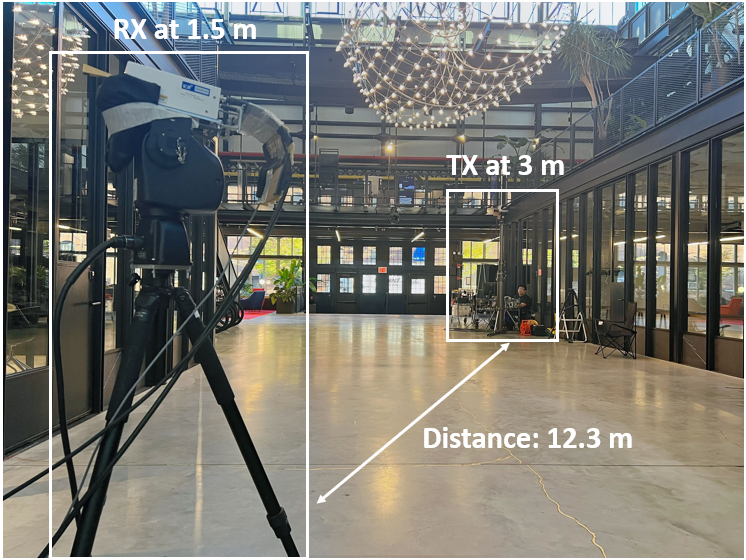}
\caption{An illustration of the measurement setup in the factory environment (TX1 and RX2). }
\label{fig:env}
\end{figure}

The 142 GHz InF channel measurements were conducted in a factory building in the Brooklyn Navy Yard, NY, in the fall of 2021. The building ($\sim$110 m L$\times$ 36 m W $\times$ 25 m H) was initially constructed in 1899 and served as a machine shop for marine engines and components, and now becomes a hub for local small manufacturing companies. As shown in Fig. \ref{fig:env}, a 6.8 m wide hallway centers in the building, and the ceiling and truss are made of steel. Manufacturing labs and conference rooms are located at two sides of the hallway, covered by large glass windows with metal frames.

We used a wideband sliding correlation-based channel sounder generating a 500 MHz pseudorandom sequence. The baseband signal was first mixed with a 7 GHz intermediate frequency (IF) signal and then up-converted to a center frequency of 142 GHz with a 1 GHz null-to-null bandwidth. The channel sounder has a two nanosecond temporal resolution for multipath components and a 152 dB maximum measurable path loss under a 5 dB signal-to-noise-ratio (SNR) constraint. The TX and RX were equipped with directional horn antennas of 27 dBi gain and an 8\degree~half-power beamwidth (HPBW). The TX and RX antennas were swept in the azimuth and elevation planes to search all the multipath components in the environment. Each azimuth sweep produced 45 directional power delay profiles (PDPs) using an 8\degree~step size. 
\begin{figure*}[h!]
\centering
\includegraphics[width=1\linewidth]{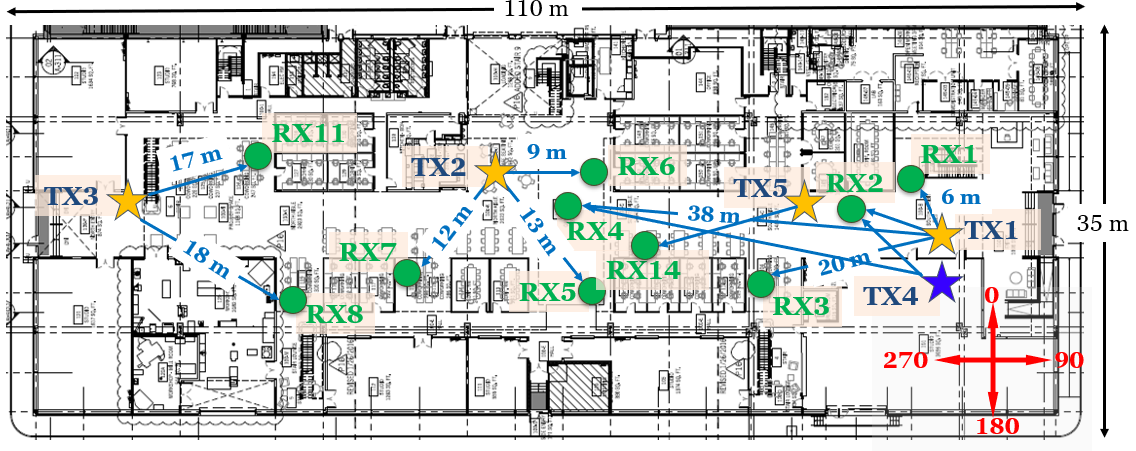}
\caption{\textcolor{black}{TX and RX locations for polarization measurements. Five TX locations are denoted as orange stars and ten RX locations are denoted as green circles, resulting in 13 TX-RX location pairs for channel measurements.} }
\label{fig:floor_plan}
\end{figure*}

\begin{figure}[h!]
\centering
\includegraphics[width=1\linewidth]{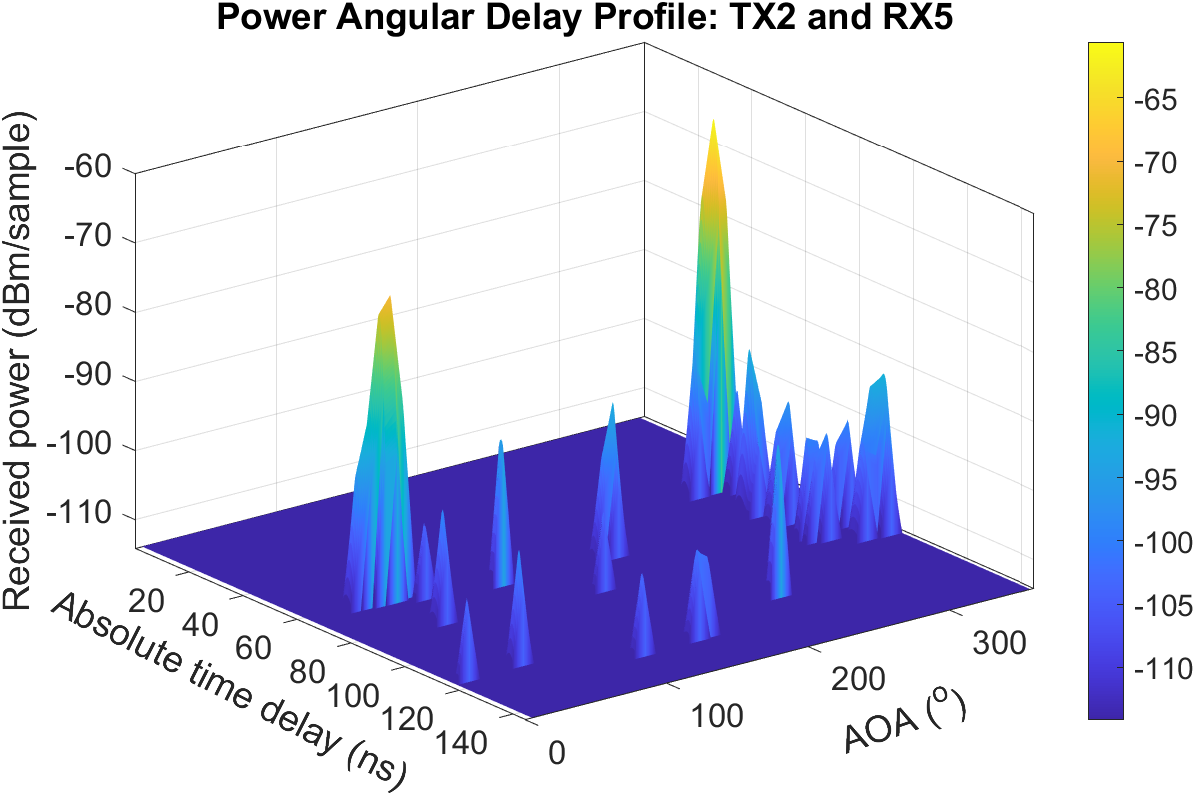}
\caption{The power angular (AOA) delay profile of TX2 and RX5. Two main arriving directions are the boresight direction and the reflection from the wall at 90\degree. The XPD values of these two directions are 28.3 dB and 22.7 dB, respectively. }
\label{fig:pdap}
\end{figure}

The TX was set at 3 m above the ground level emulating an indoor hotspot, and the RX was set at 1.5 m above the ground level as a mobile user. \textcolor{black}{Five TX locations and ten RX locations were selected for the polarization measurements to study the cross-polarization effect, resulting in 11 line-of-sight (LOS) measurement locations and two non-LOS (NLOS) measurement locations, as shown in Fig. \ref{fig:floor_plan}.} The TX-RX separation distance ranged from 6.3 m to 39.6 m. At each measurement location, the best pointing angles with the strongest received power were first found. A quick RX azimuthal power scan ($\sim$6 s) was performed for each of 45 AODs (8\degree~apart) to determine the set of AODs that have detectable powers. Then, 45 8-degree stepped rotations were conducted at the RX to record directional PDPs for each ``good'' AOD, which took about three minutes. Complete azimuth angular statistics of the double-directional channels were obtained. The vertically polarized antennas were rotated to be horizontally polarized by applying a waveguide twist between the converter output waveguide flange and the antenna. A sample power angular delay profile of TX2 and RX5 is shown in Fig. \ref{fig:pdap}. Two main arriving directions are the boresight and the reflection from the wall at 90\degree. The XPD values of these two directions are 28.3 dB and 22.7 dB, respectively. More details on the XPD will be elaborated in Section \ref{sec:pl}.

\section{Path Loss with Dual Polarization}\label{sec:pl}
This section presents the path loss data and the fitting models for both omnidirectional and directional channels. In addition, the gross XPD values of omnidirectional channels and the XPD values of different directional channels are extracted. Due to the dynamic range limit of the channel sounder, the RX3, RX4, RX8 locations only received negligible power when using the V-H polarization configuration. Thus, the path loss modeling for cross-polarization excluded these three locations and used the rest ten locations.  
\subsection{CI Path Loss Model}
We use the close-in free space reference distance (CI) path loss model with 1 m reference distance \cite{Rap13access}, as this has been proven to be superior for modeling path loss over many environments and frequencies \cite{Sun16tvt}. $\PL^\textup{CI}$ represents the path loss in dB scale, which is a function of distance and frequency:
\begin{equation}\label{eq:CI1}
\begin{split}
	\PL^\textup{CI}(f, d)[\textup{dB}]=&\textup{FSPL}(f, d_0)+10n\log_{10}\left(\frac{d}{d_0}\right)+\chi_{\sigma}^{\textup{CI}} \text{,}\\
	&\text{for}\: d\geq d_0, \;\;\text{where}\: d_0 = 1 \textup{m}
\end{split}
\end{equation}
where $n$ denotes the path loss exponent (PLE), and $\chi_{\sigma}^{\textup{CI}}$ is the shadow fading that is commonly modeled as a log-normal random variable with zero mean and $\sigma$ standard deviation in dB. $d$ is the 3-D T-R separation distance. $d_0$ is the reference distance, and $\textup{FSPL}(f, d_0)=20\log_{10}(4\pi d_0 f/c)$. The CI path loss model uses the FSPL at $d_0 = 1$ m as an anchor point and fits the measured path loss data with a straight line controlled by a single parameter $n$ (PLE) obtained via the minimum mean square error (MMSE) method. CI path loss model can be used for co-polarized or cross-polarized measurement data. 

\subsection{CIX Path Loss Model}
An extension of the basic CI path loss model for the special case of cross-polarization propagation is to add a constant attenuation factor known as the cross-polarization discrimination (XPD) factor, that best fits the measured data via an MMSE method given by:
\begin{equation}\label{eq:CIX}
\begin{split}
	\PL^\textup{CIX}(f, d)[\textup{dB}]=&\textup{FSPL}(f, d_0)+10n_{\textup{(V-V)}}\log_{10}\left(\frac{d}{d_0}\right)\\
	&+\textup{XPD} [\textup{dB}]+\chi_{\sigma}^{\textup{CIX}}
\end{split}
\end{equation}
This CI path loss with XPD (CIX) model uses the PLE measured with the co-polarized (V-V) antenna setting at identical locations and obtains the optimum XPD value via the MMSE method. As seen from (\ref{eq:CIX}), the CIX model uses the optimum XPD in dB that is added to the CI model to minimize the error between the estimated and measured cross-polarized path loss.  
\subsection{Omnidirectional Path Loss}
\begin{figure}[h!]
\centering
\includegraphics[width=1\linewidth]{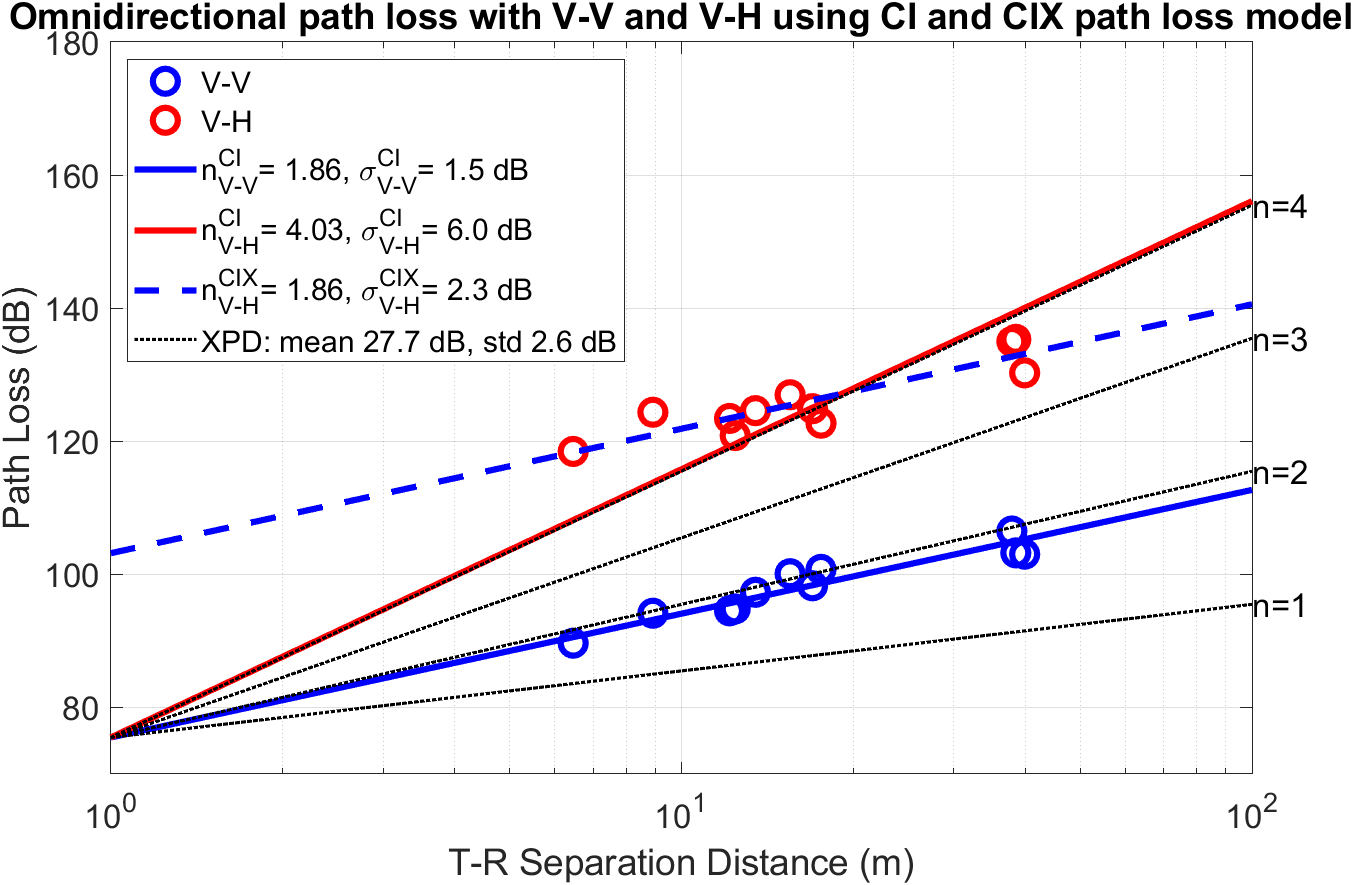}
\caption{\textcolor{black}{Omnidirectional path loss data are fitted by the CI and CIX path loss models. The gross XPD has a mean of 27.4 dB and a standard deviation of 2.3 dB.} }
\label{fig:omni_pl}
\end{figure}

The omnidirectional path loss data for V-V and V-H polarizations are shown in Fig. \ref{fig:omni_pl}. The PLE for the V-V and V-H data are 1.86 and 4.03 using the CI model, respectively. The standard deviation of the shadow fading are 1.5 dB and 6.0 dB, respectively. The CIX model using the PLE for the V-V polarization leads to the gross XPD with a mean of 27.7 dB and a standard deviation of 2.6 dB. A smaller standard deviation of the shadow fading shows that the CIX model provides a better fit than the CI model to the cross-polarization data.
\subsection{Directional Path Loss}
\begin{figure}[h!]
\centering
\includegraphics[width=1\linewidth]{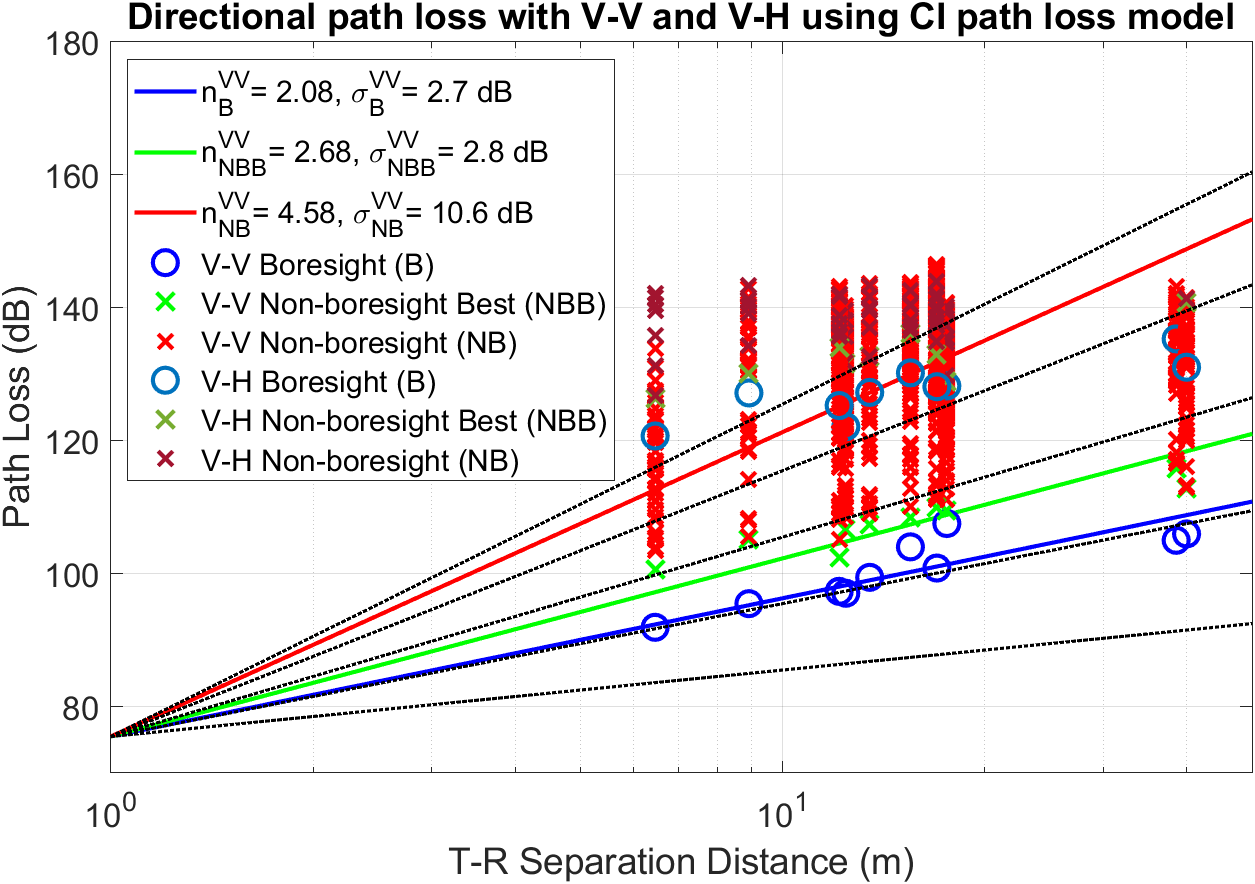}
\caption{\textcolor{black}{Directional path loss data are fitted by the CI path loss models. The NBB direction is the strongest direction in the NB direction set.}}
\label{fig:dir_pl}
\end{figure}

The directional path loss data for V-V and V-H polarizations are shown in Fig. \ref{fig:dir_pl}. Note that all the ten locations that have valid V-H path loss data are in the LOS scenario. The directions are divided into three sets: boresight (B), non-boresight best (NBB), and non-boresight (NB). The NB direction set includes all the directions that can receive a valid signal power. The NBB direction is the strongest in the NB direction set and then excluded from the NB direction set. Such classification provides important insight about the channel condition if a human or robot temporarily blocks the boresight path. Using the CI model to fit the V-V directional path loss, the resulting PLE of the B, NBB, and NB directions are 2.08, 2.68, and 4.58, respectively, indicating the best non-boresight direction is sufficiently strong to support a communication link if the boresight direction is blocked. However, an arbitrary non-boresight direction may not be usable.

\begin{figure}[h!]
	\centering
	\includegraphics[width=1\linewidth]{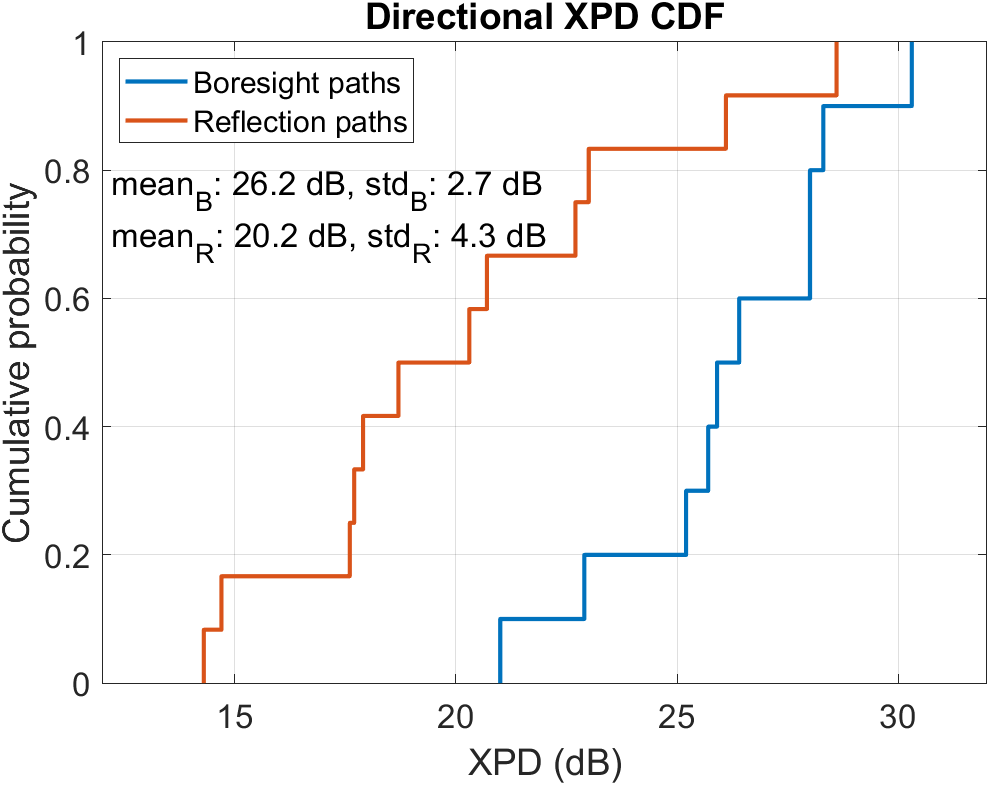}
	\caption{\textcolor{black}{Empirical cumulative distribution function (CDF) of directional XPD values of boresight paths and reflection paths.} }
	\label{fig:dir_cdf}
\end{figure}

The directions that have both valid V-V and V-H received powers are used to derive the directional XPD. We noticed that all such directions are from either the boresight path or the first-order reflection path. As shown in Fig. \ref{fig:dir_cdf}, the boresight paths have a mean XPD of 26.2 dB with a standard deviation of 2.7 dB while the reflection paths have a mean XPD of 20.2 dB with a standard deviation of 4.3 dB, showing that reflection may cause de-polarization. 

\section{Omnidirectional and Directional Delay Spread} \label{sec:dsas}
Delay spread describes the temporal dispersion of multipath channels, which is vital for the wireless system design to avoid inter-symbol interference. Note that the delay spread depends on the threshold used for data denoising. Here we applied a threshold of 20 and 30 dB below the maximum received power. The omnidirectional and directional delay spreads are given in Table \ref{tab:ds}. 
\begin{table}[]
\centering
\caption{Omnidirectional and Directional Delay Spread Statistics with 20 dB and 30 dB Threshold: Minimum, Maximum, Mean, Median, and 90 Percentile} \label{tab:ds}
\begin{tabular}{|c|c|c|c|c|c|}
	\hline
	\textbf{Delay Spread (ns)} & \textbf{Min} & \textbf{Max} & \textbf{Mean} & \textbf{Median} & \textbf{90\%} \\ \hline
	\textbf{Omni RMSDS-20 dB}  & 0.7          & 43.8         & 11.0          & 6.3             & 33.3                  \\ \hline
	\textbf{Omni RMSDS-30 dB}  & 0.7          & 66.0         & 16.0          & 10.4             & 54.4                   \\ \hline
	\textbf{Omni MDS-20 dB}    & 4.4          & 372.6        & 74.8          & 56.2            & 163.0                  \\ \hline
	\textbf{Omni MDS-30 dB}    & 7.5          & 784.8        & 157.4         & 83.9            & 490.7                 \\ \hline
	\textbf{Dir RMSDS-20 dB}   & 0.3          & 292.9        & 3.0           & 0.7             & 4.6                    \\ \hline
	\textbf{Dir RMSDS-30 dB}   & 0.3          & 292.9        & 3.0           & 0.7             & 4.6                    \\ \hline
	\textbf{Dir MDS-20 dB}     & 1.2          & 706.4        & 10.4          & 3.2             & 14.4                   \\ \hline
	\textbf{Dir MDS-30 dB}     & 1.2          & 706.4        & 10.5          & 3.2             & 14.5                   \\ \hline
\end{tabular}
\end{table} 

We choose the root mean square delay spread (RMSDS) and the maximum delay spread (MDS) to show the channel dispersion. Omnidirectional delay spreads with a 20 dB threshold are much smaller than the values with a 30 dB threshold, while the directional delay spreads do not vary much with different thresholds since a directional PDP usually contains only one to a few multipath components. Taking the omnidirectional delay spreads obtained by the 30 dB threshold, the maximum omnidirectional RMSDS and MDS are 66.0 and 784.8 ns, respectively, which shows that multipath components with significant delays may exist in the factory environment due to the metal building framework. For the directional delay spread, even though the maximum directional RMSDS and MDS are 292.9 and 706.4 ns, the 90 percentile of the directional RMSDS and MDS are only 4.6 and 14.5 ns, suggesting that most directions have very little time dispersion. 
\section{AOA and AOD Angular Spread}
\begin{figure}[]
\centering
\subfloat[2-D AOD power angular spectrum: TX1 and RX2]{\label{fig:aps1}\includegraphics[width=.8\linewidth]{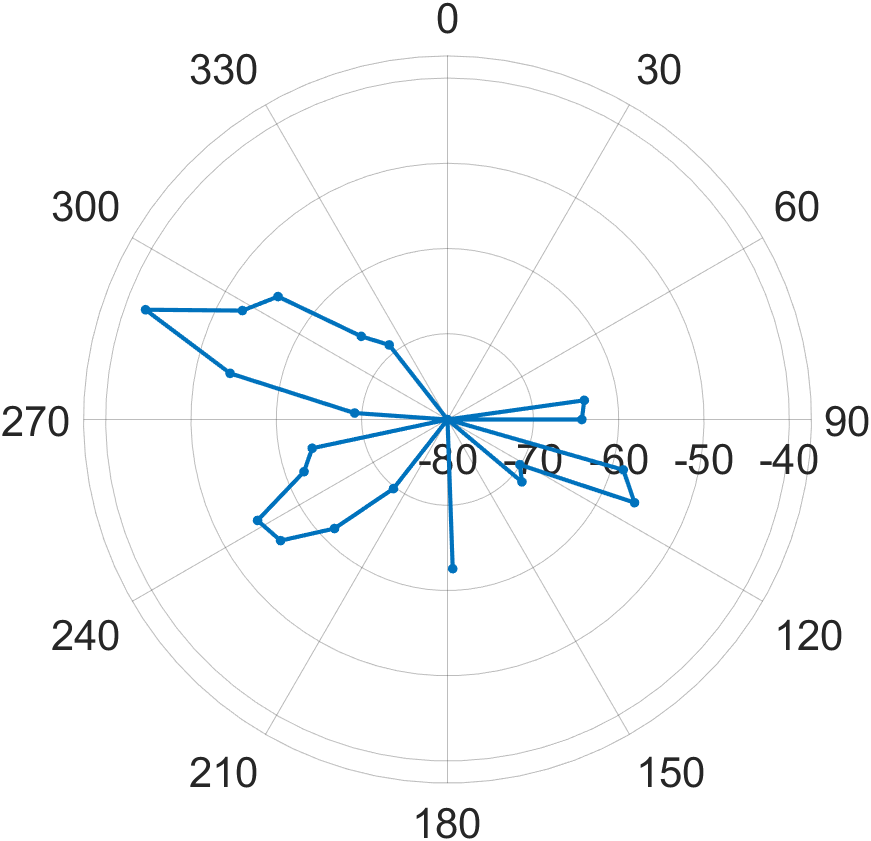}} \hspace{4mm}
\subfloat[2-D AOA power angular spectrum: TX1 and RX2]{\label{fig:aps2}\includegraphics[width=.8\linewidth]{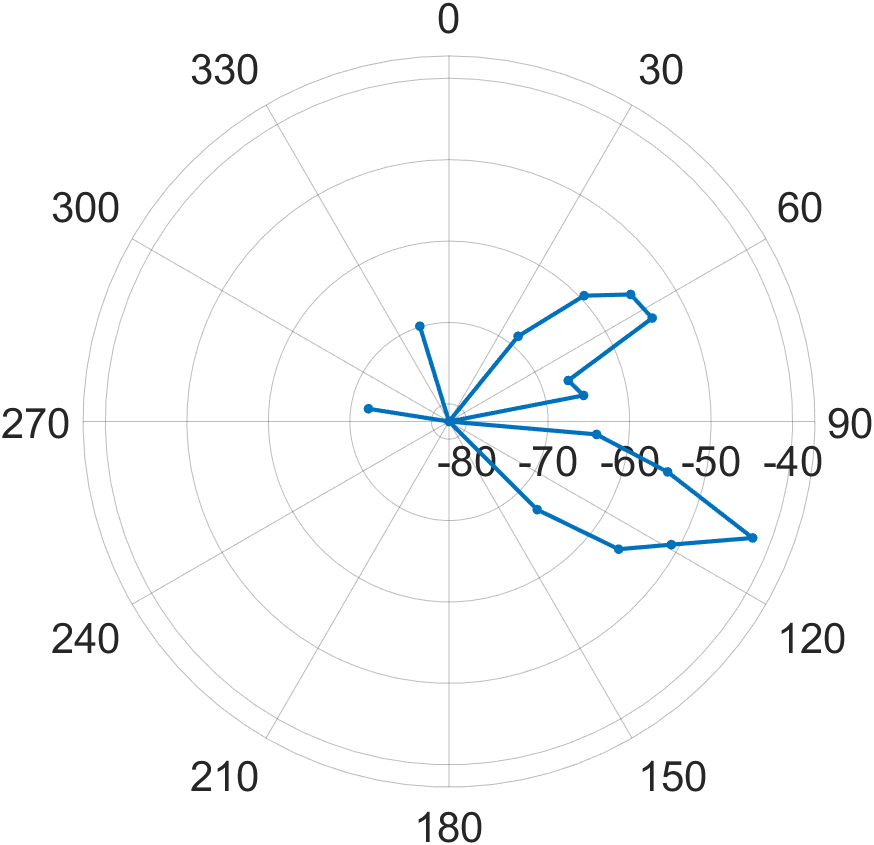}} \\ \vspace{2mm}
\caption{2-D AOD and AOA power angular spectrum for TX1 and RX2 with a 30 dB threshold.}
\label{fig:aps}
\end{figure}

The root mean square angular spread (RMSAS) describes the spatial dispersion of multipath channels, which can be computed using Appendix A-1,2 in \cite{3GPP38901r16}. Similar to the delay spread, the angular spread varies according to the threshold. Here we use a 20 or 30 dB threshold below the maximum directional power over all azimuth angles. In addition, spatial lobes are used to identify the number of distinct beams for the TX or RX. A spatial lobe represents a main direction of energy arrival or departure, which may contain multipath components over a large time window of hundreds of nanoseconds \cite{Samimi16mtt}. The sample AOD and AOA power angular spectrum for TX1 and RX2 are shown in Fig. \ref{fig:aps}, which shows five and four AOD and AOA spatial lobes. The number of AOA and AOD spatial lobes and RMSAS with 20 and 30 dB thresholds are given in Table \ref{tab:as}. 

\begin{table}[]
\centering
\caption{The number of Spatial Lobes and Angular Spread Statistics with 20 dB and 30 dB Threshold: Minimum, Maximum, Mean, Median, and 90 Percentile} \label{tab:as}
\begin{tabular}{|c|c|c|c|c|c|}
	\hline
	\textbf{Angular Spread (\degree)} & \textbf{Min} & \textbf{Max} & \textbf{Mean} & \textbf{Median} & \textbf{90\%} \\ \hline
	\textbf{\# AOA SL-20 dB}     & 1            & 4            & 2.2             & 2               & 3.2                      \\ \hline
	\textbf{\# AOA SL-30 dB}     & 1            & 5            & 3.5          & 3               & 5                    \\ \hline
	\textbf{\# AOD SL-20 dB}     & 1            & 5            & 2.2             & 2               & 3.4                     \\ \hline
	\textbf{\# AOD SL-30 dB}     & 2            & 7            & 3.5           & 3               & 7                    \\ \hline
	\textbf{AOA RMSAS-20 dB}     & 2.9          & 71.1         & 27.0          & 21            & 57.9                   \\ \hline
	\textbf{AOA RMSAS-30 dB}     & 3.2          & 72.2         & 28.1          & 22.1           & 58.1                   \\ \hline
	\textbf{AOD RMSAS-20 dB}     & 3            & 103.1        & 25.2          & 19              & 55.9                   \\ \hline
	\textbf{AOD RMSAS-30 dB}     & 4            & 103.7        & 27.4          & 21.6            & 57.2                   \\ \hline
\end{tabular}
\end{table}

The mean number of AOA spatial lobes using the 20 and 30 dB differs by 1.3, and the mean number of AOD spatial lobes using the 20 and 30 dB differs by 1.3, showing the threshold level has a notable impact on the number of spatial lobes. Also, the maximum number of AOD spatial lobes is larger than the maximum number of AOA spatial lobes. Compared to our previous work at 142 GHz in an indoor office environment \cite{Ju21jsac}, more spatial lobes are observed in the factory environment, which can be attributed to the surrounding metal structure and objects. The RMSAS show little difference using the 20 and 30 dB threshold. The mean AOA and AOD RMSAS are similar (i.e., 28.1\degree~and 27.4\degree~using a 30 dB threshold) while the maximum AOD RMSAS (103.7\degree) is much larger than the maximum AOA RMSAS (72.2\degree).

\section{Conclusion and Future Work} \label{sec:conclusion}
This paper presented recent 142 GHz channel measurements in a factory building. Five TX locations and ten RX locations were selected to investigate the channel polarization characteristics. The V-V and V-H omnidirectional path loss data were modeled using the CI and CIX models, giving a gross XPD with a mean of 27.7 dB and a standard deviation of 2.6 dB. By comparing V-V and V-H path loss in each direction, the mean XPD of the boresight path and the first-order reflection path are 26.2 dB and 20.2 dB, respectively. In addition, the best non-boresight direction produces a PLE of 2.7, which is worse than the boresight PLE of 2 but sufficient to reliably support a transmission link when the boresight link is temporarily blocked. 

The mean omnidirectional and directional RMSDS are 16.0 and 3.0 ns with a 30 dB threshold, respectively. Moreover, the multipath components with excess time delays of hundreds of nanoseconds may exist due to the metal framework of the factory building. On average, a TX and RX location pair may have 3.5 different departing and arriving directions with a 30 dB threshold. These preliminary measurements show that the channel statistics in the factory environment are different from those in the office environment. Thus, extensive channel measurements in various factory buildings are required to develop a realistic and general channel model for the InF scenario at sub-THz frequencies, which will facilitate the development of high-accuracy sensing and positioning techniques and the massive deployment of IIoT for future intelligent factories in 6G and beyond.    

\bibliographystyle{IEEEtran}
\bibliography{icc22}

\begin{thebibliography}{10}
\providecommand{\url}[1]{#1}
\csname url@samestyle\endcsname
\providecommand{\newblock}{\relax}
\providecommand{\bibinfo}[2]{#2}
\providecommand{\BIBentrySTDinterwordspacing}{\spaceskip=0pt\relax}
\providecommand{\BIBentryALTinterwordstretchfactor}{4}
\providecommand{\BIBentryALTinterwordspacing}{\spaceskip=\fontdimen2\font plus
\BIBentryALTinterwordstretchfactor\fontdimen3\font minus
  \fontdimen4\font\relax}
\providecommand{\BIBforeignlanguage}[2]{{%
\expandafter\ifx\csname l@#1\endcsname\relax
\typeout{** WARNING: IEEEtran.bst: No hyphenation pattern has been}%
\typeout{** loaded for the language `#1'. Using the pattern for}%
\typeout{** the default language instead.}%
\else
\language=\csname l@#1\endcsname
\fi
#2}}
\providecommand{\BIBdecl}{\relax}
\BIBdecl

\bibitem{Chen18access}
B.~Chen, J.~Wan, L.~Shu, P.~Li, M.~Mukherjee, and B.~Yin, ``Smart factory of
  {Industry} 4.0: Key technologies, application case, and challenges,''
  \emph{IEEE Access}, vol.~6, pp. 6505--6519, 2018.

\bibitem{Tao19tii}
F.~Tao, H.~Zhang, A.~Liu, and A.~Y.~C. Nee, ``Digital twin in industry:
  State-of-the-art,'' \emph{IEEE Transactions on Industrial Informatics},
  vol.~15, no.~4, pp. 2405--2415, 2019.

\bibitem{Cheng18a}
J.~Cheng, W.~Chen, F.~Tao, and C.-L. Lin, ``Industrial {IoT} in {5G}
  environment towards smart manufacturing,'' \emph{Journal of Industrial
  Information Integration}, vol.~10, pp. 10--19, 2018.

\bibitem{Rap19access}
T.~S. {Rappaport} \emph{et~al.}, ``Wireless communications and applications
  above 100 {GHz}: Opportunities and challenges for {6G} and beyond,''
  \emph{IEEE Access}, vol.~7, pp. 78\,729--78\,757, June 2019.

\bibitem{FCC19}
{Federal Communications Commission}, ``Spectrum horizons,'' \emph{First Report
  and Order – ET Docket 18-21}, March 2019.

\bibitem{3GPP38901r16}
3GPP, ``Technical specification group radio access network; study on channel
  model for frequencies from 0.5 to 100 {GHz (Release 16)},'' TR 38.901
  V16.0.0, October 2019.

\bibitem{Maltsev10AwpLetter}
A.~{Maltsev}, E.~{Perahia}, R.~{Maslennikov}, A.~{Sevastyanov}, A.~{Lomayev},
  and A.~{Khoryaev}, ``Impact of polarization characteristics on {60-GHz}
  indoor radio communication systems,'' \emph{IEEE Antennas and Wireless
  Propagation Letters}, vol.~9, pp. 413--416, 2010.

\bibitem{Sun14CommMag}
S.~Sun,  \emph{et~al.}, ``{MIMO} for millimeter-wave wireless communications:
  beamforming, spatial multiplexing, or both?'' \emph{IEEE Communications
  Magazine}, vol.~52, no.~12, pp. 110--121, 2014.

\bibitem{80211ad10}
A.~Maltsev \emph{et~al.}, ``Channel models for 60 {GHz WLAN} systems,'' doc.:
  IEEE 802.11-09/0334r8, May 2010.

\bibitem{Ju21jsac}
S.~Ju, Y.~Xing, O.~Kanhere, and T.~S. Rappaport, ``Millimeter wave and
  sub-{Terahertz} spatial statistical channel model for an indoor office
  building,'' \emph{IEEE Journal on Selected Areas in Communications}, vol.~39,
  no.~6, pp. 1561--1575, 2021.

\bibitem{Xing21b}
Y.~Xing, T.~S. Rappaport, and A.~Ghosh, ``Millimeter wave and {sub-THz} indoor
  radio propagation channel measurements, models, and comparisons in an office
  environment,'' \emph{IEEE Comm. Letters}, pp. 1--1, 2021.

\bibitem{Rap13access}
T.~S. Rappaport \emph{et~al.}, ``Millimeter wave mobile communications for {5G}
  cellular: It will work!'' \emph{IEEE Access}, vol.~1, pp. 335--349, May 2013.

\bibitem{Sun16tvt}
S.~{Sun} \emph{et~al.}, ``Investigation of prediction accuracy, sensitivity,
  and parameter stability of large-scale propagation path loss models for {5G}
  wireless communications,'' \emph{IEEE Transactions on Vehicular Technology},
  vol.~65, no.~5, pp. 2843--2860, May 2016.

\bibitem{Samimi16mtt}
M.~K. {Samimi} and T.~S. {Rappaport}, ``{3-D} millimeter-wave statistical
  channel model for {5G} wireless system design,'' \emph{IEEE Transactions on
  Microwave Theory and Techniques}, vol.~64, no.~7, pp. 2207--2225, July 2016.

\end{thebibliography}

\end{document}